\documentclass[a4paper,12pt]{article}

\usepackage{hyperref}
\usepackage{braket}
\usepackage{amssymb}
\usepackage{amsmath}
\usepackage{amsthm}
\usepackage{amscd}
\usepackage{fullpage}

\theoremstyle{definition}

\theoremstyle{remark}

\newcommand{\abs}[1]{\lvert#1\rvert}

\begin{document}

\title{Hidden Messenger from Quantum Geometry:
Towards Information Conservation in Quantum Gravity}

\author{Xiao-Kan Guo\footnote {E-mail: kankuohsiao@whu.edu.cn }\\
Wuhan Institute of Physics and Mathematics,\\ Chinese Academy of Sciences, Wuhan 430071, China; \\
University of  Chinese Academy of Sciences, Beijing 100049, China\\~\\
Qing-yu Cai\\
Wuhan Institute of Physics and Mathematics,\\ Chinese Academy of Sciences, Wuhan 430071, China; \\
Department of Physics, Florida International University,\\
 11200 SW 8th Street, CP 204
Miami, FL 33199, USA
}

\date{\today}

\maketitle

\begin{abstract}
The back reactions of Hawking radiation allow nontrivial correlations between consecutive Hawking quanta, which gives a possible way of resolving the paradox of black hole information loss known as the hidden messenger method.
In a recent work of Ma {\it et al} [arXiv:1711.10704], this method is enhanced by a general derivation using small deviations of the states of Hawking quanta off canonical typicality. In this paper, we use this typicality argument to study the effects of generic back reactions on the quantum geometries described by spin network states, and discuss the viability of entropy conservation in loop quantum gravity. We find that such back reactions lead to small area deformations of quantum geometries including those of quantum black holes. This shows that the hidden-messenger method is still viable in loop quantum gravity, which is a first step towards resolving the paradox of black hole information loss in quantum gravity.
\end{abstract}
\newpage

\section{Introduction}
The paradox of black hole information loss due to the thermal nature of Hawking radiation \cite{Haw76} has refused a  final resolution for more than forty years. The attempts to understand this paradox  have brought about both numerous stimulating ideas and many concrete theoretical models for black holes and quantum gravity. See e.g. \cite{BC11} for  reviews of various calculable proposals to resolve this paradox.

Since the original arguments leading to the paradox given in \cite{Haw76} (and of course the derivation of Hawking radiation) are semiclassical, the first thought is to solve this problem also at the semiclassical level. This is what the majority of literatures focus on, the reasons of which are that, on the one hand, for an asymptotic observer the quantum effects can be effectively approximated by  semiclassical descriptions, and on the other hand, for a long period of time there are no full-fledged quantum theories of gravity at our disposal. The most famous semiclassical prescription is perhaps the black hole complementarity \cite{STU94}. But the arguments of AMPS \cite{AMPS} add further complications to this proposal by ``fusing" the theory of black holes with quantum information theory. See \cite{Harlow} for a review.

There is, however, a simple information-theoretical proposal known as the {\it hidden messengers} \cite{ZCYZ09}, based on non-thermal radiations derived from the semiclassical tunneling method \cite{PW00}. The problem of the hidden-messenger approach is that it is too simple: the argument for information conservation given in \cite{ZCYZ09} is just the chain rule of conditional entropy without enough details of the physical processes, and the derivation of non-thermal radiation is restricted to the tunneling method which only calculates the tunneling rate of Hawking quanta. The tunneling method has been enumeratively applied to all kinds of black hole solutions in recent years, but surprisingly for linear dilaton black holes the Hawking radiation calculated by the tunneling method is {\it thermal} \cite{Sak}. In this case, the hidden-messenger argument holds only if one further takes the quantum-gravitational corrections into account. This poses the question of whether the hidden messengers are semicalssical or fully quantum,  which  requires the knowledge of quantum gravity.

There are, of course, attempts to study the hidden messengers beyond the tunneling picture. A recent example  is \cite{HC17} where the arguments are based on entropic gravity, which is in a sense the easiest way out, because  the information conservation arguments via hidden messengers only deal with the black hole entropy.
Recently, Ma {\it et al} \cite{MCDS17} present a simple way to derive the hidden-messenger arguments via (deviations from) canonical typcality. This approach enables us to generalize the arguments to explicit theories of quantum gravity. In this work, we  consider the existence of hidden messengers in the spin network states of loop quantum gravity via canonical typicality. As a proof of principle, we show that in the kinematical  Hilbert space of spin-network intertwiners, a small deformation from the typical spin network state  studied in  the recent works of Anza and Chirco \cite{AC} allows a local hidden messenger to inform such a deformation. This is qualitively true both for black-hole-like configurations and for ordinary quantum geoemtries, which therefore shows in loop quantum gravity the viability of the entropy conservation by back reactions. 

 In Sec. \ref{s2}, we first critically review the hidden-messenger arguments for the information conservation of black holes, and then single out the simple derivation via typicality. In Sec. \ref{s3}, after recalling the typical spin network state, we perform a second trace on this typical state and show the existence of hidden messengers both by considering the form of probability and by evaluating the mutual information. It turns out that the hidden messengers correspond to the closure defects of the quantum geometry as the area deformations of the quantum polyhedra. Further remarks are presented in Sec. \ref{conc}.

\section{Hidden messenger in Hawking radiation: general arguments}\label{s2}
The so-called hidden messenger in Hawking radiation means the correlation between two consecutive Hawking quanta. The existence of such correlations is simply due to the fact that when we take into account the energy conservation during the Hawking radiance, the emission rate of Hawking quanta is proportional to $e^{\Delta S}$ where $\Delta S$ is the change in the black hole entropy. The formula in the tunelling method are not necessarily needed for the purpose of information consevation.  These facts already exist implicitly in \cite{guo}.
To see this, let us consider the following schematic derivation:

Firstly, the emission rate of Hawking radiation is generically of the form $\Gamma\propto e^{\Delta S}$, where $S$ is the black hole entropy which could be the Bekenstein-Hawking entropy or others with corrections.  Now suppose the prefactor of the {\it probability} $P(\propto\int\Gamma dt)$ has been taken care of and set to 1, then we can write the probability $P(E_1)\equiv e^{S(M-E_1)-S(M)}$ for the first emission with energy $E_1$ from the black hole with mass $M$, and for the second emission as the conditional probability $P(E_2|E_1)\equiv e^{S(M-E_1-E_2)-S(M-E_1)}$. It is easy to show that $P(E_1,E_2)=P(E_1)P(E_2|E_1)=P(E_1+E_2)$, and by the chain rule one has $P(E_1,...,E_n)=P(\sum_{i=1}^nE_i)$ which through exponetiation entails the so-called conservation of entropy.

Till now we have not mentioned  the hidden messengers, the correlations between Hawking quanta. This is because the form $e^{\Delta S}$  already includes the back reactions. To reveal those hidden messengers, one simply considers the correlation between the probabilities of two consecutive Hawking quanta
\begin{equation}
C(E_1,E_2)=\ln P(E_1+E_2)-\ln[P(E_1)P(E_2)]=\ln P(E_2|E_1)-\ln P(E_2)\neq0.
\end{equation}
This is not surprising in view of the back reactions and the energy conservation. The picture of the hidden-messenger proposal thus returns to the unitary picture of  burning paper. A nice way to understand this unitary picture is the so-called Bohr-like approach to black hole physics \cite{bohr}: semiclassically a black hole is the gravitational analog of Bohr's hydrogen atom with Hawking radiations as gravitational ``electrons", and its unitary time evolution is effectively described by a time dependent Schr\"odinger equation.

As we can see now, the hidden-messenger arguments are only information-theoretical and not dependent on the physical models involved. (Even the discrete nature of the spectrum of Hawking radiation is a model-independent quantum feature \cite{LC16}). We can understand the word ``messenger" as the definable conditional probability due to the effects of the back reactions of Hawking radiation. But these conditional probabilities are also definable for burning papers. So if we forget the physical models, the hidden-messenger arguments will say nothing about the black hole physics. In this sense, these messengers are still ``hidden". 
Notice that the relation $\Gamma\propto e^{\Delta S}$ can be obtained from several different appraoches \cite{PW00,MCDS17,MP00}. One can  work with those explicit models and interpret the back reaction of Hawking radiation on geometries as physical messengers. But there are still some complications:
\begin{enumerate}
\item The arguments in support of the information conservation are physically rather vague. If one interprets the black entropy as the entanglement entropy of entangled states across the horizon, then the recovery of information is the recovery of the whole entanglement (or entangled pairs) which cannot be obtained soly by those outgoing quanta. On the other hand, if one interprets the black hole entropy as the Boltzmann entropy of the microstates, then what need to be recovered are these configurations of microstates instead of their entropies. (The attempted interpretation given in \cite{int11} is a kind of circular logic that one recovers the initial configurations identified from the final entropy formula.)
The point is that entropies only mathematically measure the information, and what are physical are those quantum states carrying information.  It is therefore necessary to return to physical models to reconsider the hidden messengers themselves.
\item Recent progress on the semiclassical Einstein equations shows that the back reactions of Hawking radiation will deform the near horizon geometry into a  geometry without event horizon \cite{KMY}. This challenges many conventional calculation methods such as the tunelling method, although the picture of back reaction will probably still be valid.
\end{enumerate}

In order to forward the hidden messengers, working in new frameworks will be helpful. So let us recall the approach via canonical typicality recently presented in \cite{MCDS17}. Consider the total Hilbert space $\mathcal{H}_U=\mathcal{H}_S\otimes\mathcal{H}_E$ where $\mathcal{H}_S$ stands for the Hilbert space of the small system and $\mathcal{H}_E$ for the Hilbert space of the large environment. The system could be a black hole, and in that case one can impose the global (in the sense that they are macroscopic) constriant $\mathcal{R}$ as the energy (or other hairs) conservation during the evaporation process. This constraint picks a subspace $\mathcal{H}_{\mathcal{R}}$ out of $\mathcal{H}_U$. By canonical typicality \cite{typ}, for any $\ket{\phi}\in\mathcal{H}_{\mathcal{R}}$, the reduced state of the system is closed to the canonical state
\begin{equation}
\rho_S(\phi)=\text{Tr}_E(\ket{\phi}\bra{\phi})=\text{Tr}_E({\bf1}_{\mathcal{R}}/d_{\mathcal{R}})
\end{equation}
where ${\bf1}_{\mathcal{R}}$ is the projector onto  $\mathcal{H}_{\mathcal{R}}$ and $d_{\mathcal{R}}=\dim\mathcal{H}_{\mathcal{R}}$. Now suppose a black hole described in $\mathcal{H}_S$ has energy $E_b$ and the Hawking radiation in $\mathcal{H}_E$ has energy $E_r$, i.e. $\mathcal{H}_{\mathcal{R}}=\mathcal{H}_b\otimes\mathcal{H}_r$, then the constraint is simply $E_b+E_r=E$ with $E$ being the total energy for states in $\mathcal{H}_{\mathcal{R}}$. Consequently, the state of Hawking radiation is
\begin{equation}\label{3}
\rho_r=\text{Tr}_b\rho_S(\phi)=\sum_r\frac{d_{\mathcal{R}}-d_r}{d_{\mathcal{R}}}\ket{r}\bra{r}=\sum_re^{\Delta S_b}\ket{r}\bra{r},\quad \ket{r}\in\mathcal{H}_r,
\end{equation}
where $S_b$ is the Boltzmann entropy of the black hole. The probability  thus retains the form $e^{\Delta S}$.

Again, to fully appreciate this approach, we need to know the details of these Hilbert spaces. In the following section, we will work with a concrete model in loop quantum gravity. The goal is to see the physical contents of these messengers in concrete quantum gravity models on the premise that the entropy conservation holds.
\section{Kinematical messenger in spin network states}\label{s3}
The spin network states form a basis of the kinematical Hilbert space of loop quantum gravity. They also encodes the discrete quantum geoemtries. For a rigorous treatment one is referred to the monographs \cite{LQG}. Pointedly stated, the  kinematical Hilbert space of loop quantum gravity is defined on a background independent graph $\gamma$ with $E$ edges and $V$ vertices. To each edge $e$ is assigned a half-integer $j_e$ which labels an irreducible representation of SU(2), the gauge group of loop quantum gravity. The corresponding representation space $V^{j_e}$ is of dimension $(2j_e+1)$. Since the edges intersect at the vertices, to each vertex $v$ is assigned an intertwiner $\mathcal{I}_v:\otimes_{e\ni v}V^{j_e}\rightarrow\mathbb{C}$, the invariant tensor. For fixed $j_e$ the intertwiners form a Hilbert space $\mathcal{H}_{v,j_e}$ with intertwiners $\ket{\mathcal{I}_v}\in\otimes_v\mathcal{H}_{v,j_e}$. Then the spin network states $\ket{\gamma,\{j_e\},\{\mathcal{I}_v\}}$ form a basis of the kinematical Hilbert space 
\begin{equation}
\mathcal{H}_\gamma=L^2(\text{SU(2)}^E/\text{SU(2)}^V,d\mu_{\text{Haar}})=\bigoplus_{j_e}\bigotimes_v\mathcal{H}_{v,j_e}.
\end{equation}
The spin network basis is in fact an orthonormal basis. Indeed, the projections of  intertwiners onto the $\text{SU(2)}^E/\text{SU(2)}^V$ connection representation define the spin network wave functions
\begin{equation}\label{55}
\psi_{\{j_e\},\{\mathcal{I}_v\}}[g_e]=\braket{g_e|\mathcal{I}_v}=\text{Tr}\bigotimes_eD^{j_e}(g_e)\bigotimes_v\mathcal{I}_v,\quad g_e\in\text{SU(2)}
\end{equation}
where the Wigner $D$-matrices have orthogonality. By choosing the basis of intertwiner Hilbert spaces, one gets the orthonormality.
The actions of geometric operators  on the spin network states will give quantized discrete spectra. For instance, the area operator acts like 
\begin{equation}\label{5}
\hat{A}\ket{\gamma,\{j_e\},\{\mathcal{I}_v\}}=\ell^2_pj_e\ket{\gamma,\{j_e\},\{\mathcal{I}_v\}},\quad\ell_p:~\text{Planck length}.
\end{equation}

With these in mind, let us turn to the typicality in spin network states \cite{AC}. Consider the total Hilbert space of $N$ edges each spin of which is bounded by a constant $J\gg1$. Suppose the system has $k$ edges and the environment has $N-k$ edges with $N\gg k$. That is, we consider the  Hilbert space
\begin{equation}
\mathcal{H}=\bigoplus_{j_e\leqslant J}\bigotimes_{i=1}^NV^{j_i}=\bigoplus_{j_e\leqslant J}\Bigl\{\bigoplus_{q,r}\Bigl[\Bigl(V^q\otimes V^r\Bigr)\otimes\Bigl(d_q^{\{j_S\}}\otimes d_r^{\{j_E\}}\Bigr)\Bigr]\Bigr\}\equiv\mathcal{H}_S\otimes\mathcal{H}_E,
\end{equation}
where we have used the fact that the tensor product of irreducible representations can be decomposed into the direct sum 
\begin{equation}
\bigotimes_{i}V^{j_i}=\bigoplus_qV^q\otimes d_q^{\{j_i\}},\quad d_q^{\{j_i\}}:~\text{degeneracy spaces of spin-$q$}.
\end{equation}

Next, the SU(2)-invariant $N$-valent intertwiners  form a Hilbert space $\mathcal{H}_N^{(J)}$ with constant total area $J(=\sum_e j_e)$ \cite{FR10}. (This is the inrreducible representation space of U(N), which is the starting point of the spinorial/twistorial representation of loop quantum gravity.) We can therefore impose the global constraint by fixing a constant total area $J_0$ such that $\mathcal{H}_\mathcal{R}$ becomes
\begin{equation}\label{9}
\mathcal{H}_\mathcal{R}=\mathcal{H}_N^{(J_0)}=\bigoplus_{J_S\leqslant J_0}\bigoplus_{\abs{\vec{J}_S}}\Bigl\{\Bigl(\bigoplus_{\{j_S\}}^{J_S}d_k^{\{j_S\},\abs{\vec{J}_S}}\Bigr)\otimes\Bigl(\bigoplus_{\{j_E\}}^{J_0-J_S}d_{N-k}^{\{j_E\},\abs{\vec{J}_S}}\Bigr)\Bigr\}
\end{equation}
where the $V$'s are contracted by the single $N$-valent intertwiner due to the SU(2) invariance, $\abs{\vec{J}_S}$ is the total angular momentum of the system as the closure defect (meaning the breaking of the closure constraint), and the sum over $\{j\}$ keeps the total area. As we can see from \eqref{9} this is  semidecoupled in the sense that an intertwiner is splitted into two intertwiners one of which is in the system $S$ and the other in the environment $E$ with an internal edge connecting them. The edges in $S$ and $E$ are coupled within $S$ and $E$ respectively, and the internal edge couples $S$ and $E$ carrying $\abs{\vec{J}}_S$. The semidecoupled basis of $\mathcal{H}_{\mathcal{R}}$ is then
\begin{align}
\sum_{M_S}\frac{(-1)^{\abs{\vec{J}}_S+M_S}}{\sqrt{2\abs{\vec{J}}_S+1}}&\ket{\{j_S,j_E\};\sigma_S,\eta_E,\abs{\vec{J}}_S,M_S;\abs{\vec{J}}_S,-M_S}=\nonumber\\
=&\sum_{M_S}\frac{(-1)^{\abs{\vec{J}}_S+M_S}}{\sqrt{2\abs{\vec{J}}_S+1}}\ket{\{j_S\};\sigma_S;\abs{\vec{J}}_S,M_S}_S\otimes\ket{\{j_E\};\eta_E;\abs{\vec{J}}_S,-M_S}_E\label{10}
\end{align}
where $M_S$ is the projection of $\abs{\vec{J}}_S$ along the ``$z$-axis", $\eta_E,\sigma_S$ are the SU(2) recoupling quantum numbers, and the kets in \eqref{10} are coupled bases in the angular momentum representation. Then the projector onto  $\mathcal{H}_{\mathcal{R}}$ can be expanded in terms of the semidecoupled basis,
\begin{align}
{\bf1}_{\mathcal{R}}=\sum_{\{j_S,j_E\}}^{J_0}&\sum_{\sigma_S,\eta_E}\sum_{\abs{\vec{J}}_S,M_S,M_S'}\frac{(-1)^{M_S+M'_S}}{d_{\abs{\vec{J}}_S}}\times\nonumber\\
&\times\ket{\{j_S,j_E\};\sigma_S,\eta_E,\abs{\vec{J}}_S,M_S;\abs{\vec{J}}_S,-M_S}\bra{\{j_S,j_E\};\sigma_S,\eta_E,\abs{\vec{J}}_S,M'_S;\abs{\vec{J}}_S,-M'_S}
\end{align}
where $d_{\abs{\vec{J}}_S}=2\abs{\vec{J}}_S+1$.
The dimension $d_{\mathcal{R}}$ is given by the dimension of the irreducible representation of U(N) with the highest weight $[J_0,J_0,0,...]$,
\begin{equation}
d_{\mathcal{R}}=\frac{1}{J_0+1}\begin{pmatrix}N+J_0-1\\J_0\end{pmatrix}\begin{pmatrix}N+J_0-2\\J_0\end{pmatrix}.
\end{equation}

The canonical state can be readily calculated as
\begin{align}
\rho_S=&\text{Tr}_E({\bf1}_{\mathcal{R}}/d_{\mathcal{R}})=\nonumber\\
=&\sum_{J_S\leqslant J_0/2}\sum_{\sigma_S,\abs{\vec{J}}_S,M_S}\sum_{\{j_S\}}^{J_S}\frac{D_{N-k}(\abs{\vec{J}}_S,J_0-J_S)}{d_{\abs{\vec{J}}_S}d_{\mathcal{R}}}\ket{\{j_S\},\sigma_S,\abs{\vec{J}}_S,M_S}_S\bra{\{j_S\},\sigma_S,\abs{\vec{J}}_S,M_S},
\end{align}
where the canonical weight
\begin{align}
D_{N-k}&(\abs{\vec{J}}_S,J_0-J_S)=\sum_{\{j_E\}}^{J_0-J_S}\sum_{\eta_E}\sum_{\abs{\vec{J}}_E}\delta_{\abs{\vec{J}}_S,\abs{\vec{J}}_E}=\nonumber\\
=&\frac{d_{\abs{\vec{J}}_S}}{\abs{\vec{J}}_S+J_0-J_S+1}\begin{pmatrix}N-k+\abs{\vec{J}}_S+J_0-J_S-1\\\abs{\vec{J}}_S+J_0-J_S\end{pmatrix}\begin{pmatrix}N-k-\abs{\vec{J}}_S+J_0-J_S-2\\-\abs{\vec{J}}_S+J_0-J_S\end{pmatrix}.\label{14}
\end{align}
In \eqref{14} we have used the dimension formula of the U(N) representation with the highest weight $[l_1,l_2,0,...]$ where $l_1-l_2=2\abs{\vec{J}}_S,l_1+l_2=2J_E$. The dependence of $\abs{\vec{J}}_S$ in \eqref{14} shows the local correlation structure preserving the  global constraints which are broken by the splitting of the intertwiner. This already indicates the existence of the locally hidden messengers.

The typicality requires the trace distance $D(\rho,\rho_S)$ between an arbitrary pure state $\rho$ in $\mathcal{H}_{\mathcal{R}}$ and the canonical state $\rho_S$ is very small. This restricts the range of  parameters such as $J_0$ and $N$. As has been shown in \cite{AC}, in the large area limit $J_0\gg N$ with $J_0/N$ finite suitable for semiclassical black holes, there is a wide range of parameters where the typicality holds. In this regime, the von Neumann entropy of the system's canonical state has the leading order contribution
\begin{equation}\label{15}
S_{J_0\gg N\gg1}\sim2k\Bigl(1+\log\bigl(\frac{J_0}{N}\bigr)\Bigr)+....=\ln\Bigl(e(\frac{J_0}{N})^{1/\ln2}\Bigr)^{2k}+...
\end{equation}
This entropy is extensive in $k$, the number of edges or dually the elementary area patches in the system $S$. This is consistent with the quantum picture of black hole horizon proposed in \cite{LT06} since it is in the form of a logarithm of a dimension of some effective degeneracy space.
On the other hand, the large $N$ regime is of interest for quantum gravity. In this case the system's von Neumann entropy is shown in \cite{AC} to be
\begin{equation}\label{16}
S_{N\gg J_0\gg1}\sim\Bigl(1+\log\bigl(\frac{N-k}{J_0}\bigr)\Bigr)\braket{2J_S}+....\equiv\frac{1}{T}E+...
\end{equation}
where the last expression comes from the Schwinger boson representation of SU(2) spin networks \cite{FR10} where the energy of a pair of oscillators on an edge equals the the spin on that edge. 

In \cite{LT06}, the evaporation of a black hole is described kinematically as the detachment of a pair of (spin-$\frac{1}{2}$) qubits  (living on a single area patch) from the horizon. By detechment it means that $j=0$ on the edge connecting this pair to the rest of spin networks, or equivalently there is no edge between these two parties. In the present case, we consider the detachment of area patches as decoupling some edges in the system $S$. 
In other words, we further decompse $\mathcal{H}_S$ as if there is only $(k-r)$ edges in $S$, 
\begin{equation}
\mathcal{H}_S=\bigoplus_{\abs{\vec{J}}_S}\Bigl\{\Bigl(\bigoplus_{\{j_S\}}^{J_S-J_d}d_k^{\{j_S\},\abs{\vec{J}}_S}\Bigr)\bigotimes_{\{j_d\}}^{J_d} V^{j_d}\Bigr\}\equiv\mathcal{H}_b\otimes\mathcal{H}_d,
\end{equation}
where $j_d$ is the spin on the decoupled edge. Notice that in this decomposition the SU(2) invariance is broken outside $\mathcal{H}_b$, because the Hawking radiation, if any, should be a dynamical effect and hence not in the kinematical Hilbert space. Meanwhile, the breaking of the global constraint of energy conservation will induce an internal edge carrying closure defect $\abs{\vec{J}}_d$.
 The coupled basis can then be expanded with the help of CG coefficients,
\begin{equation}
\ket{\{j_S\},\sigma_S,\abs{\vec{J}}_S,M}_S=\sum_{M}\sum_{m_d} C^{\abs{\vec{J}}_d,M-m_d,j_d,m_d}_{\abs{\vec{J}}_S,M}\ket{\abs{\vec{J}}_d,M-m_d,j_d,m_d}.
\end{equation}
We can now trace over the remaining coupled edge, as in \eqref{3}, so as to get the state of the ``Hawking radiation". Since the Hilbert space $\mathcal{H}_b$ still carries the U($k-r$) representation, the trace gives rise to another canonical weight function $D$. By the splitting of sum
\[\sum_{\{j_S\}}^{J_S}\rightarrow\sum_{J_d\leqslant J_S/2}\sum_{\{j_d\}}^{J_d}\sum_{\{j_b\}}^{J_S-J_d},\]
we have
\begin{align}\label{19}
\rho_d=\text{Tr}_b\rho_S=&\sum_{J_S\leqslant J_0/2}\sum_{\abs{\vec{J}}_S,M}\frac{D_{N-k}(\abs{\vec{J}}_S,J_0-J_S)}{d_{\abs{\vec{J}}_S}d_{\mathcal{R}}}\times\nonumber\\
&\times\sum_{J_d\leqslant J_S/2}\sum_{\{j_d\}}^{J_d}\sum_{\abs{\vec{J}_d},m_d}C^2(M,j_d,m_d){D_{k-r}(\abs{\vec{J}}_d,J_S-J_d)}\ket{j_d,m_d}\bra{j_d,m_d}
\end{align}
where the $C$ are CG coefficients. Again, the closure defect $\abs{\vec{J}}_d$ retains the global constraint of energy conservation. 

The total weight in \eqref{19} is expected to be proportional to $e^{\Delta S}$. To see this, we first notice that when $J_S\gg J_d$ (or equivalently $k\gg r$), the deviation from the canonical state $\rho_S$ is small and hence $\rho_b$ will remain typical. The entropy of the remaining system $b$ is therefore maximal and proportional to $\ln D_{k-r}(\abs{\vec{J}}_d,J_S-J_d)$, which means that the angular momentum basis of the intertwiner Hilbert space provides a complete description of the configurations of a ``black hole". We can then  take the weight in $\rho_d$ as the deviation from the canonical weight  near that in $\rho_S$ and repeat the approximation of the dimension function $D$ performed in \cite{AC} to extract the form $e^{\Delta S}$. Also notice that we are dealing with the spin network states describing arbitrary quantum geometries instead of those of black holes. So in order to describe black holes, we should restrict ourselves to the regime where the entropy of $b$ obeys the area law even though it can be expressed as $\ln\Omega_{\text{eff}}$ for some effective dimension $\Omega_{\text{eff}}$ (cf. also the second paper of \cite{AC}). In the typicality regime, either the  entropy \eqref{15} or \eqref{16} follows the area law: $S_{J_0\gg N\gg1}$ is extensive in the number of area patches and respectively the $\braket{2J_S}\sim E=2A$ in $S_{N\gg J_0\gg1}$ are quantized areas. But only \eqref{15} allows a Boltzmann form. Hence, the reduced state \eqref{19} can  be expressed schematically as
\begin{equation}
\rho_r\propto\sum_r\frac{\Omega_{b}}{\Omega_{S}}\ket{r}\bra{r}=\sum_re^{\Delta S}\ket{r}\bra{r}
\end{equation}
where $1/\Omega_S$ comes from the term $D_{N-k}(\abs{\vec{J}}_S,J_0-J_S)/(d_{\abs{\vec{J}}_S}d_{\mathcal{R}})$ as a consequence of typicality, and the form of $\Omega_b$ can be roughly seen from the Stirling approximation,
\begin{equation}\label{21}
\log\Bigl({D_{k-r}(\abs{\vec{J}}_d,J_S-J_d)}\Bigr)\sim\log\Bigl(\frac{eJ_S}{k-r}\Bigr)^{2(k-r)}+...
\end{equation}
where we have assumed $J_S\gg k\gg1$, the large area limit. Different from the entanglement entropy in \eqref{15} and \eqref{16}, we interpret $\ln\Omega_b$ as the Boltzmann entropy of the ``black hole" $b$. Therefore,  the probability of ``Hawking radiation" is proportional to $e^{\Delta S}$, at least  formally in the large area limit.

The form $e^{\Delta S}$ allows us to apply the arguments for entropy conservation of the last section. Since the back reaction are ubiquitous, we expect these arguments still apply in non-black-hole regimes. For instance, the  system-bath coupling in ordinary quantum systems is a similar structure \cite{XLLS14}. A crude way of seeing this is to consider the mutual information between two consecutive parts of ``radiation", say $r_1$ and $r_2$. By the definition of the $D$ functions, we have
\begin{equation}\label{22}
\log\frac{D_{N-k}D_{k-r_1-r_2}}{D_{N-k}D_{k-r_1}D_{N-k}D_{k-r_2}}=\log\frac{D_{k-r_1-r_2}}{D_{k-r_1}D_{k-r_2}}-\log D_{N-k}\neq0.
\end{equation}
In fact, the term $\log D_{N-k}$ appears in the von Neumann entropy of $\rho_S$, and hence we can consider a term like $\log D_{k-r_1}$ as proportional to the entropy of an effective system (obtained from the first trace) with $k-r_1$ edges as long as $k\gg r_1$. This way \eqref{22} holds in that, as a matter of principle, $S_{k-r_1-r_2}-S_{k-r_1}-S_{k-r_2}-S_k\neq0$. As an example, consider the large number-of-edge  limit $N\gg J_0,k\gg J_d$  where the system's entropy is given by \eqref{16}. If we further assume $r_1,r_2\ll k$, the inverse temperatures of ``radiations" can be taken as identical so that \eqref{22} means 
\begin{equation}
\beta_r(\braket{2(J_{d_1}+J_{d_2})}-\braket{2J_{d_1}}-\braket{2J_{d_2}})-\text{const.}\braket{2J_{S}}\sim-E_S\neq0.
\end{equation}
We thus qualitively see  that the mutual information is  non-vanishing.

In particular, since the large number-of-edge  limit the entropy \eqref{16} is consistent with the Schwinger representation of SU(2) spin networks, we can generate the closure defects by the deformation operators \cite{Liv14}. Indeed, in the Schwinger boson representation \cite{FR10}, an edge in a spin network is replaced by a pair of quantum harmonic oscillators with annihilation operatos ($a_i,b_j$)  (with the $\hat{\cdot}$ omitted) which form the basis of a quantized spinor. Without going into the details of the spinor representation, we can consider the generators of $\mathfrak{u}(N)$, in view of the U(N) representation of intertwiners,
\begin{equation}
E_{ij}=a_i^\dagger a_j+b_i^\dagger b_j.
\end{equation}
Then the Casimir of this $\mathfrak{u}(N)$ is the deformation operator
\begin{equation}
\mathcal{E}^2=\frac{1}{2}\sum_{i,j}E_{ij}E_{ji}=\frac{1}{2}E_{ii}^2+\frac{1}{2}E_{ii}(N-2)+\mathcal{C}^2\equiv A^2+\abs{J}^2
\end{equation}
where  $A=\frac{1}{2}E_{ii}$ is the area of the boundary, $\mathcal{C}^2$  is the $\mathfrak{su}(2)$ Casimir operator, and $\abs{J}^2$ is the closure defect. Likewise, for the cross product $F_{ij}=a_ib_i-b_ia_j$, one has $F^2=A^2-\abs{J}^2$. Now that the total energy of the oscillators are $E=\sum_iE_{ii}$, we see that the small deformation by the closure defects will give rise to a change in the entropy  given in \eqref{16}. So even in the non-black-hole case, by the canonical typicality the small deformation in energy at the density matrix of a U(N) intertwiner's edge still resume the coefficients $e^{-\beta \Delta\braket{E}}\sim e^{\Delta S}$.

Notice that in \cite{LT06} the entanglement between a detached area patch and the rest spin networks is zero. On the contrary, here detaching an edge results in a closure defect on the virtual link, and hence the mutual information is nonzero. The correlation shown in the mutual information is in fact the sum of the intertwiner entanglement between two splitted intertwiners, the spin entanglement on the internal virtual edge as is distinguished in \cite{Liv17}, and possibly other types of quantum correlations. In \cite{Liv17} the spin entanglement on a spin network edge is argued to be unphysical since it breaks the gauge invariance. Here, as has been emphasized in \cite{AC}, it is exactly the breaking of global constraints when one traces out some subsystem that induces such local structures of correlation. This spin entanglement on the virtual internal edge, with entnaglement entropy $S=\ln(2\abs{\vec{J}}_d+1)$, plays the role of a messenger who preserves of the global constraint of energy conservation. In this sense, we have qualitively recovered the arguments about energy conservation and back reactions in the last section.


\section{Concluding remarks}\label{conc}
In conclusion, we have studied in this paper the viability of the hidden-messenger approach  to the paradox of black hole information loss in the quantum gravitational descriptions of black holes. The technique we exploit is the small deviations from the canonical typicality given by Ma {\it et al} recently, instead of the original one using the tunneling method. An emphasis has been put on the  back reactions of Hawking radiation and their effects on deforming quantum geometries. We applied this technique to the kinematical Hilbert space of spin-netwrok intertwiners and found that in a black-hole-like regime the probability of  Hawking radiation formally allows the existence of hidden messengers. 
This is also true for ordinary quantum geometries as can be seen from the nonvanishing mutual information between consecutive radiated states. Therefore, the hidden-messenger appraoch to the problem of black hole information loss is implementable in loop quantum gravity.
Moreover, we identified the messengers  as the small deformation in quantum geometry, which gives a concrete meaning to them.

It is desirable to describe Hawking radiation in terms of the full dynamics of quantum gravity. Pranzetti \cite{Pran} has applied Thiemann's solution to the Hamiltonian constriant of loop quantum gravity to the near horizon region, and used grand canonical ensemble of the spin-network punctures to describe the changes in the spins of puncturing edges. Two edges $p,q$ of the bulk spin netwrok are changed into $p\rightarrow p-\frac{1}{2},q\rightarrow q\pm\frac{1}{2}$ that piercing the horizon with an spin-$\frac{1}{2}$ edge inside the horizon, so that the intensity of the (energetic) spectrum lines are changed to have a thermal envelope. From viewpoint of this paper, the changes of spins in the edges piercing of the horizon resulting in effectively new punctures should cause small area deformations of the local area patches, which in principle resumes the hidden-messenger arguments of entropy conservation. The deviation from thermal envelope will be a possible signature of such effects, as the Bose-Hubbard model of  intertwiner dynamics \cite{FR17} may suggest. (In a recent paper \cite{MCS18}, the mutual information appeared in the entropy conservation arguments is called ``dark" in the sense that they are unobservable just as dark energy. In view of the possible phenomenological effects of discrete quantum geometries, their assertions are too hasty to be made.)

Can we conclude that the  black hole information paradox is then resolved in quantum gravity? Unfortunately, we still can not. As we have mentioned above, the hidden-messenger arguments only ensure the entropy conservation. To resolve the the paradox, the conservation of information in the mathematical sense is not enough, although it is of course necessary. By saying information, we are really interested in what the information is about. In the case of black holes, we are interested in, for example, what the infalling and asymptotic observers will see. The entropy conservation does not say much about these aspects. Although we have shown that these messengers are deformations of quantum geometry, this is only a proof of principle. In order to find more operational criteria, one still has much work to do. 

Finally, let us also remark on the intertwiner and spin entanglement: As we have mentioned, in \cite{Liv17} the spin   entanglement on a spin network edge is argued to be unphysical due to its breaking of the gauge invariance. But the statistical origin of the black hole entropy is usually explained in loop quantum gravity as due to the topological defects coming from the punctures of spin-network edges on the horizon (see e.g. \cite{LQG}). It is an old story \cite{Car95} that such a horizon  breaks the  gauge symmetry and generates topological defects on the horizon. So there seems to be a contradiction   as pointed out in \cite{Liv17}.
The hidden messegers revealed in this paper provides another way to understand this issue: the black hole horizon with spin networks' punctures indeed breaks the gauge symmetry, but it is so locally on the puncturing edges which, as messengers, inform the preservation of the global gauge invariance. Since these messengers arise in the physical picture like Hawking radiation,  the question arises as to whether the origin of black hole entropy, if it can be described by the topological defects on the horizon, is dynamical.

\section*{Acknowledgements} 
We thank Sumanta Chakraborty for helpful comments. Financial support from National Natural Science Foundation of China under Grant Nos. 11725524, 61471356
and 11674089 is gratefully acknowledged.

\bibliographystyle{amsalpha}

\begin{thebibliography}{99}
\bibitem{Haw76} S. W. Hawking, Breakdown of predictability in gravitational collapse, Phys. Rev. D {\bf14},
2460 (1976).
\bibitem{BC11} V. Balasubramanian, B. Czech, Quantitative approaches to information recovery from black holes, Class. Quantum Grav. {\bf28}, 163001 (2011); S. Chakraborty, K. Lochan, Black holes: Eliminating information or illuminating new physics? Universe, {\bf3}, 55 (2017).
\bibitem{STU94} L. Susskind, L. Thorlacius, J. Uglum, The stretched horizon and black hole complementarity, Phys. Rev. D {\bf48}, 3743 (1994).
\bibitem{AMPS} A. Almheiri, D. Marolf, J. Polchinski, J. Sully, Black holes: complementarity or firewalls?,
JHEP02(2013)062; A. Almheiri, D. Marolf, J. Polchinski, D. Stanford, J. Sully, An apologia for firewalls, JHEP09(2013)018.
\bibitem{Harlow} D. Harlow, Jerusalem lectures on black holes and quantum information, Rev. Mod. Phys. {\bf88}, 015002 (2016).
\bibitem{ZCYZ09} B.-c. Zhang, Q.-y. Cai, L. You, M.-s. Zhan, Hidden messenger revealed in Hawking radiation: a resolution to the paradox of black hole information loss, Phys. Lett. B {\bf675}, 98 (2009).
\bibitem{PW00} M. K. Parikh, F. Wilczek, Hawking radiation as tunneling, Phys. Rev. Lett. {\bf85}, 5042 (2000);  M. Angheben,  M. Nadalini,  L. Vanzo, S. Zerbini, Hawking radiation as tunneling for extremal and rotating black holes, JHEP05(2005)014.
\bibitem{Sak} I. Sakalli, M. Halilsoy, H. Pasaoglu, Entropy conservation of linear dilaton black holes in quantum corrected Hawking radiation, Int. J. Theor. Phys. {\bf50}, 3212 (2011); I. Sakalli, A. Ovgun, Uninformed Hawking radiation, EPL {\bf110}, 10008 (2015).

\bibitem{HC17} D.-s. He, Q.-y. Cai, Gravitational correlation, black hole entropy, and information conservation,  Sci. China-Phys. Mech. Astron. {\bf60}, 040011 (2017).
\bibitem{MCDS17} Y.-H. Ma, Q.-y. Cai, H. Dong, C.-P. Sun, Non-thermal radiation of black hole off canonical typicality, arXiv:1711.10704v3.
\bibitem{AC} F. Anza, G. Chirco, Typicality in spin-network states of quantum geometry, Phys. Rev. D {\bf94}, 084047 (2016); Fate of the hoop conjecture in quantum gravity, Phys. Rev. Lett. {\bf119}, 231301 (2017).

\bibitem{guo} H. Dong, Q.-y. Cai, X.-F. Liu, C.-P. Sun, One hair postulate for Hawking radiation as tunneling process, Commun. Theor. Phys. {\bf61}, 289 (2014); X.-K. Guo, Q.-y. Cai, Information recovery with Hawking radiation from dynamical horizons, Int. J. Theor. Phys. {\bf53}, 2980 (2014).
\bibitem{bohr} C. Corda, Precise model of Hawking radiation from the tunnelling mechanism, Class. Quantum Grav. {\bf32}, 195007 (2015); Time dependent Schrödinger equation for black hole evaporation: no information loss, Ann. Phys. {\bf353}, 71 (2015).

\bibitem{LC16} K. Lochan, S. Chakraborty, Discrete quantum spectrum of black holes, Phys. Lett. B {\bf755}, 37 (2016); Decoding infrared imprints of quantum origins of black holes, arXiv:1711.10660.
\bibitem{MP00} S. Massar, R. Parentani, How the change in horizon area drives black hole evaporation, Nucl. Phys. B {\bf575}, 333 (2000); R. Zhao, L.-C. Zhang, S.-Q. Hu, A new method for exploring the black hole Hawking radiation--the quantum statistical method, Acta Physica Sinica {\bf55}, 3898 (2006);
Y.-P. Zhang, Q. Dai, W.-B. Liu, Unthermal Hawking radiation from a general stationary black hole, Commun. Theor. Phys.  {\bf49}, 379 (2008).
\bibitem{int11} B.-c. Zhang, Q.-y. Cai, M.-s. Zhan, L. You, An interpretation for the entropy of a black hole, Gen. Relativ. Gravit. {\bf43}, 797 (2011).
\bibitem{KMY} H. Kawai, Y. Mastuo, Y. Yokokura, A self-consistent model of black hole evaporation, Int. J. Mod. Phys. A {\bf28}, 1350050 (2013); V. Baccetti, R. B. Mann, D. R. Terno, Do event horizons exist?, Int. J. Mod. Phys. D {\bf26}, 743008 (2017).
\bibitem{typ}   S. Goldstein, J. Leibowitz, R. Tumulka,  N. Zanghi, Canonical typicality, Phys. Rev. Lett. {\bf96}, 050403 (2006);  S. Popescu, A. J. Short,  A. Winter, Entanglement and the foundations of statistical mechanics, Nat. Phys. {\bf2}, 754 (2006).
\bibitem{LQG} T. Thiemann, {\it Modern Canonical Quantum General Relativity}, Cambridge University Press, 2007;  C. Rovelli, F. Vidotto, {\it Covariant Loop Quantum Gravity}, Cambridge University Press, 2014.
\bibitem{FR10}  F. Girelli, E. R. Livine, Reconstructing quantum geometry from quantum information: spin networks as harmonic oscillators, Class. Quantum Grav. {\bf22}, 3295 (2005); L. Freidel, E. R. Livine, The fine structure of SU(2) intertwiners from U(N) representations, J. Math. Phys. {\bf51}, 082502 (2010).
\bibitem{LT06} E. R. Livine, D. R. Terno, Quantum black holes: entropy and entanglement on the horizon, Nucl. Phys. B {\bf741}, 131 (2006).
\bibitem{XLLS14} D.-Z. Xu, S.-W. Li, X.-F. Liu, C.-P. Sun, Noncanonical statistics of a finite quantum system with non-negligible system-bath coupling, Phys. Rev. E {\bf90}, 062125 (2014).
\bibitem{Liv14} E. R. Livine, Deformation operators of spin networks and coarse-graining, Class. Quantum Grav. {\bf31}, 075004 (2014).
\bibitem{Liv17} E. R. Livine, Intertwiner entanglement on spin networks, Phys. Rev. D {\bf97}, 026009 (2018).

\bibitem{Pran} D. Pranzetti, Radiation from quantum weakly dynamical horizons in loop quantum gravity, Phys. Rev. Lett. {\bf109}, 011301 (2012); Dynamical evaporation of quantum horizons,  Class. Quantum Grav. {\bf30}, 165004 (2013).
\bibitem{FR17} A. Feller, E. R. Livine, Quantum surface and intertwiner dynamics in loop quantum gravity, Phys. Rev. D {\bf95}, 124038 (2017).
\bibitem{MCS18} Y.-H. Ma, J.-F. Chen, C.-P. Sun, Dark information of black hole radiation raised by dark energy,  Nucl. Phys. B {\bf931}, 418 (2018).
\bibitem{Car95} S. Carlip, Statistical mechanics of the (2+1)-dimensional black holes, Phys. Rev. D
{\bf51}, 632 (1995).
\end{thebibliography}

\end{document}